\documentclass[final,5p,times,twocolumn]{elsarticle}

\usepackage{amssymb}
\usepackage{booktabs}
\usepackage{multirow}
\usepackage{graphicx}
\biboptions{numbers,sort&compress}
\usepackage[table]{xcolor}

\usepackage{amsmath}
\usepackage{xcolor}

\journal{Pattern Recognition Letters}

\begin{document}

\begin{frontmatter}



\title{Fairness-Aware Partial-label Domain Adaptation for \\ Voice Classification of Parkinson’s and ALS} 





\author[inst1,inst2]{Arianna Francesconi}
\ead{arianna.francesconi@epfl.ch}

\author[inst1]{Zhixiang Dai}
\ead{zhixiang.dai@epfl.ch}

\author[inst1]{Arthur Stefano Moscheni}
\ead{arthur.moscheni@epfl.ch}

\author[inst1]{Himesh Morgan Perera Kanattage}
\ead{himesh.kanattage@epfl.ch}

\author[inst3]{Donato Cappetta}
\ead{d.cappetta@eustema.it}

\author[inst3]{Fabio Rebecchi}
\ead{f.rebecchi.guest@eustema.it}

\author[inst2,inst4]{Paolo Soda\corref{cor1}}
\ead{p.soda@unicampus.it}

\author[inst2]{Valerio Guarrasi\fnref{equal}}
\ead{valerio.guarrasi@unicampus.it}

\author[inst5]{Rosa Sicilia\fnref{equal}}
\ead{rosa.sicilia@unicamillus.org}

\author[inst1]{Mary-Anne Hartley\fnref{equal}}
\ead{mary-anne.hartley@epfl.ch}

\cortext[cor1]{Corresponding author.}
\fntext[equal]{These authors contributed equally to this work.}


\affiliation[inst1]{organization={School of Computer and Communication Sciences, EPFL (\'Ecole polytechnique f\'ed\'erale de Lausanne)},
      city={Lausanne},
      country={Switzerland}}

\affiliation[inst2]{organization={Unit of Artificial Intelligence and Computer Systems, Universit\`a Campus Bio-Medico di Roma},
      city={Rome},
      country={Italy}}

\affiliation[inst3]{organization={Eustema S.p.A., Research and Development Centre},
      city={Naples},
      country={Italy}}

\affiliation[inst5]{organization={UniCamillus-Saint Camillus International University of Health Sciences},
      city={Rome},
      country={Italy}}

\affiliation[inst4]{organization={Department of Diagnostics and Intervention, Radiation Physics, Biomedical Engineering, Umeå University},
      city={Umeå},
      country={Sweden}}

\begin{abstract}
Voice-based digital biomarkers can enable scalable, non-invasive screening and monitoring of Parkinson’s disease (PD) and Amyotrophic Lateral Sclerosis (ALS). However, models trained on one cohort or device often fail on new acquisition settings due to cross-device and cross-cohort domain shift. This challenge is amplified in real-world scenarios with partial-label mismatch, where datasets may contain different disease labels and only partially overlap in class space. In addition, voice-based models may exploit demographic cues, raising concerns about gender-related unfairness, particularly when deployed across heterogeneous cohorts.
To tackle these challenges, we propose a hybrid framework for unified three-class (healthy/PD/ALS) cross-domain voice classification from partially overlapping cohorts. The method combines style-based domain generalization with conditional adversarial alignment tailored to partial-label settings, reducing negative transfer. An additional adversarial gender branch promotes gender-invariant representations.
We conduct a comprehensive evaluation across four heterogeneous sustained-vowel datasets, spanning distinct acquisition settings and devices, under both domain generalization and unsupervised domain adaptation protocols. The proposed approach is compared against twelve state-of-the-art machine learning and deep learning methods, and further evaluated through three targeted ablations, providing the first cross-cohort benchmark and end-to-end domain-adaptive framework for unified healthy/PD/ALS voice classification under partial-label mismatch and fairness constraints. 
Across all experimental settings, our method consistently achieves the best external generalization over the considered evaluation metrics, while maintaining reduced gender disparities. Notably, no competing method shows statistically significant gains in external performance.
\end{abstract}




\end{frontmatter}



\section{Introduction}
Voice-based digital biomarkers are emerging as a scalable and non-invasive tool for the detection and monitoring of neurological disorders such as Parkinson’s disease (PD) and Amyotrophic Lateral Sclerosis (ALS)~\cite{gope2020raw}. In both conditions, vocal impairments reflect neuromuscular alterations, often appearing early and progressing alongside functional decline. Sustained phonation tasks are therefore particularly suitable for pattern recognition approaches supporting early screening and remote assessment. 
Recent machine learning (ML) and deep learning (DL) methods have shown promising results in binary disease detection (e.g., PD vs.\ healthy or ALS vs.\ healthy), sometimes achieving high accuracy on small, homogeneous cohorts~\cite{sedigh2025voice,ibarra2023towards}.
In contrast, multi-disease voice classification has received very limited attention. To the best of our knowledge, only one study has explored ternary HC/PD/ALS classification within a single dataset~\cite{gope2020raw}, whilst no prior work has investigated unified HC/PD/ALS voice classification across heterogeneous datasets or addressed cross-cohort generalization, domain adaptation, or label-space mismatch.

Beyond task-level limitations, real-world deployment of voice-based diagnostic models remains challenging.
Models must generalize across substantial cross-device and cross-cohort variability, arising from heterogeneous microphones, recording conditions, and population characteristics, which can severely degrade performance~\cite{rahmatallah2025pre,sedigh2025voice}.
At the same time, voice signals encode demographic information, and models may inadvertently exploit gender-related cues, resulting in systematic performance disparities across demographic groups~\cite{yang2024deconstructing,yang2024limits}.

To address generalization under domain shift, two main paradigms are typically adopted: domain generalization (DG) and unsupervised domain adaptation (UDA).
In both scenarios, \textit{source} domains refer to datasets available during training, whilst \textit{target} domains correspond to new acquisition settings encountered at deployment. 
DG focuses on learning domain-invariant representations that generalize to unseen target domains without any access to target data during training. 
In contrast, UDA assumes the availability of unlabelled target samples and exploits them to align source and target data distributions, relying solely on source-domain labels.
Within this context, a particularly challenging and underexplored scenario is \textit{partial-label domain shift}, where training data originate from multiple source domains whose label spaces only partially overlap. For instance, one source domain may contain samples with label space $\mathcal{Y}_s^{(1)} = \{y_1, y_2\}$, 
whereas another source provides $\mathcal{Y}_s^{(2)} = \{y_2, y_3\}$, 
and no single dataset covers the complete label set $\{y_1, y_2, y_3\}$.
Although source label spaces are known, their mismatch with the (unknown) target label space can induce negative transfer from source-only classes~\cite{cao2018partial}.

In this work, we address a novel cross-domain voice-based classification task, which jointly involves multi-disease (HC/PD/ALS) discrimination and the fusion of multiple cohorts with partially overlapping label spaces.
We investigate both DG and UDA settings and introduce \textit{FairPDA} (Fairness-aware Partial-label Domain Adaptation), an end-to-end framework that combines style augmentation, partial-label adversarial alignment, and adversarial gender debiasing, encouraging representations that are robust to domain shift and invariant to speaker gender~\cite{yang2023adversarial}.
Our main contributions are:
\begin{itemize}
    \vspace{-2.5mm}
    \item \textit{Cross-domain evaluation}: a systematic evaluation of sustained-vowel PD and ALS classifiers across heterogeneous datasets, highlighting the performance degradation observed under cross-device and cross-cohort shift. To the best of our knowledge, this is the first cross-cohort evaluation setup for ternary HC/PD/ALS voice classification across heterogeneous datasets.
    \vspace{-2.5mm}
    \item \textit{Hybrid partial-label adaptation}: a DL approach combining MixStyle-based domain generalization with adversarial partial-label UDA, aiming to improve transfer under both domain shift and label-space mismatch.
    \vspace{-2.5mm}
    \item \textit{Fairness-aware learning}: the integration of adversarial gender debiasing into a domain-adaptive voice classification model, with a systematic evaluation of performance and fairness trade-offs.
\end{itemize}

\section{Related Work}
This section reviews prior work along three directions: voice-based detection of PD and ALS, domain generalization/adaptation in medical AI, and fairness-aware learning. We cite representative and widely adopted approaches, prioritizing relevance to cross-cohort voice modelling, PD/ALS applications when available, and validated effectiveness in speech or healthcare settings. 

\textbf{Voice-Based Detection of PD and ALS.}
Voice analysis has long been investigated for the detection of neurodegenerative diseases, using both hand-crafted acoustic features and learned representations. Early studies in PD detection primarily relied on clinically motivated descriptors such as Mel-frequency cepstral coefficients (MFCCs) and standardized paralinguistic feature sets (e.g., eGeMAPSv02~\cite{eyben2015geneva}), combined with classical ML classifiers and evaluated mostly within-dataset~\cite{ibarra2023towards}. More recent work adopts DL architectures operating on time--frequency representations or raw audio, showing improved robustness to recording and linguistic variability~\cite{sedigh2025voice,rahmatallah2025pre}.
In the ALS domain, DL-based approaches have been explored for binary ALS versus HC classification, typically on limited datasets~\cite{gope2020raw}.
For instance, Gope et al.~\cite{gope2020raw} proposed a raw-waveform deep model (1D-CNN followed by BLSTM layers) and evaluated discrimination between PD vs.\ HC and ALS vs.\ HC. However, this study does not investigate cross-cohort evaluation, domain adaptation, or fairness; therefore, its behaviour under heterogeneous acquisition settings and cohort shift remains unclear.
In line with these limitations, several studies report marked performance degradation when models are evaluated on unseen cohorts~\cite{ibarra2023towards,sedigh2025voice}, suggesting sensitivity to dataset-specific artefacts rather than disease-related vocal markers.

\textbf{Domain Adaptation in Medical AI.}
To address performance degradation induced by domain shift, DG and UDA techniques have been increasingly explored in medical AI~\cite{guan2021domain}.
Given the breadth of this field, we focus on representative approaches that explicitly modify or regularize intermediate feature representations to mitigate domain shift, aligning with the methodological scope of our framework.
Among DG approaches, MixStyle~\cite{zhou2021domain} simulates domain variability by mixing feature statistics across samples, improving cross-domain robustness~\cite{zhou2024MixStyle}. Although originally introduced in computer vision, its feature-level perturbation strategy is applicable to speech-based cross-cohort modelling.
Conversely, UDA methods such as Domain-Adversarial Neural Networks (DANN)~\cite{ganin2016domain} learn domain-invariant features via adversarial training and have been applied to voice-based PD detection~\cite{ibarra2023towards}, whereas no analogous adaptation studies have been reported for ALS voice analysis. 
Conditional variants such as CDAN~\cite{long2018conditional} improve adversarial alignment by conditioning on class predictions, while statistical methods like CORAL~\cite{sun2016deep} reduce discrepancies between feature distributions; although primarily evaluated in computer vision, both are transferable to speech-based domain adaptation settings.
Most UDA methods assume shared label space between source and target domains, limiting applicability under label-space mismatch~\cite{guan2021domain}. Although partial-label domain adaptation has been explored in computer vision~\cite{cao2018partial}, it has not been systematically investigated in medical voice analysis, and, to the best of our knowledge, no prior work addresses partial-label domain shifts in multi-disease PD/ALS settings.

\textbf{Fairness and Adversarial Debiasing.}
Algorithmic fairness has become a central concern in healthcare AI, as models may exhibit systematic performance disparities across demographic groups~\cite{yang2024limits}. In speech-based applications, gender-related differences have been shown to induce statistically significant prediction gaps~\cite{yang2024deconstructing}. To our knowledge, fairness has not yet been systematically evaluated for voice-based PD or ALS detection.

Fairness metrics such as demographic parity and equalized odds quantify disparities in prediction rates and error rates across groups~\cite{xu2022algorithmic}. Existing bias mitigation strategies span pre-processing techniques (e.g., re-sampling data), in-processing methods (e.g., adversarial learning), and post-processing adjustments to model outputs. Among in-processing strategies, adversarial debiasing is a flexible approach in speech-related tasks that promotes invariance to sensitive attributes by introducing a competing adversary~\cite{yang2023adversarial,yang2024deconstructing}.

In summary, prior work reveals three gaps: (i)~ternary HC/PD/ALS classification under cross-cohort evaluation remains largely unexplored; (ii)~robustness to domain shift with partially overlapping label spaces has not been addressed in PD/ALS voice modelling; and (iii)~gender fairness has not been systematically studied in this context. These gaps serve as the foundation of the proposed end-to-end framework that jointly targets domain shift, partial-label mismatch, and demographic bias. 


\section{Datasets and cohort definition}
We consider four publicly available sustained–phonation voice datasets collected in heterogeneous clinical and acquisition settings: the \textit{mPower} mobile study and the \textit{UAMS} cohort, both including PD and HC participants, and the \textit{VOC-ALS} and \textit{Minsk} datasets, both comprising ALS and HC cohorts. All datasets include sustained vowel phonations acquired using different devices and protocols, resulting in heterogeneity in channel bandwidth, sampling rate, and acquisition protocols. 
The \textit{mPower}~\cite{bot2016mpower} and \textit{VOC-ALS}~\cite{dubbioso2024voice} datasets were collected via different custom smartphone applications for remote voice acquisition, sampled at 44.1\,kHz and 8\,kHz, respectively. The \textit{Minsk} dataset~\cite{vashkevich2021classification} was recorded in a controlled clinical setting using smartphones with standard headsets (44.1\,kHz), whilst the \textit{UAMS} dataset~\cite{prior2023voice} comprises telephone-channel clinical recordings (i.e., acquired over standard telephony and thus limited to telephone bandwidth) at 8\,kHz, rather than app-based smartphone recordings.

The \textit{mPower} cohort includes multiple recordings per participant (17.63 on average), with phonations lasting approximately $10.04 \pm 0.02$\,s.
In contrast, \textit{UAMS}, \textit{VOC-ALS}, and \textit{Minsk} provide exactly one recording per participant, with variable durations ($3.30 \pm 1.30$\,s, $14.71 \pm 7.29$\,s, and $3.83 \pm 1.66$\,s, respectively).
To harmonize phonatory content across datasets, we restrict the analysis to sustained vowel /a/ phonations, the only vowel shared across all cohorts, in line with prior studies~\cite{gope2020raw,rahmatallah2025pre}. Moreover, since gender is the only demographic attribute common to all datasets, fairness analysis is restricted to male and female groups.

\begin{table}[!t]
\centering
\caption{Patient distribution per dataset, class, and gender. Values are reported as number of patients (\#) and percentage (\%). Gender percentages are computed within each class and total percentages within each dataset.}
\resizebox{\columnwidth}{!}{%
\begin{tabular}{l|l|c|c|c}
\hline
Dataset & Class & Male (\#,\%) & Female (\#,\%) & Total (\#,\%) \\
\hline
\multirow{2}{*}{mPower}
 & HC  & 751 (86.4\%) & 118 (13.6\%) & 869 (66.6\%) \\
 & PD  & 311 (71.5\%) & 124 (28.5\%) & 435 (33.4\%) \\
\hline
\multirow{2}{*}{VOC-ALS}
 & HC  & 31 (63.3\%) & 18 (36.7\%) & 49 (33.8\%) \\
 & ALS & 62 (64.6\%) & 34 (35.4\%) & 96 (66.2\%) \\
\hline
\multirow{2}{*}{UAMS}
 & HC  & 14 (38.9\%) & 22 (61.1\%) & 36 (49.3\%) \\
 & PD  & 19 (51.4\%) & 18 (48.6\%) & 37 (50.7\%) \\
\hline
\multirow{2}{*}{Minsk}
 & HC  & 13 (39.4\%) & 20 (60.6\%) & 33 (54.1\%) \\
 & ALS & 14 (50.0\%) & 14 (50.0\%) & 28 (45.9\%) \\
\hline
\end{tabular}}
\label{tab:patient_counts}
\end{table}

\paragraph{Cohort definition and filtering}
To ensure data quality and remove sources of noise unrelated to the learning task, we applied a consistent set of inclusion and exclusion criteria. Only subjects with valid diagnostic labels, age information, and reported gender were retained. Following prior work~\cite{rahmatallah2025pre}, participants reporting comorbid conditions known to affect speech production were excluded (the complete list of excluded comorbidities is provided in Section~A.1 of the Supplementary Material).
Additional dataset-specific filters were applied to mitigate medication-related confounds, excluding recordings with missing medication metadata or acquired under non-standard medication conditions (PD subjects recorded \textit{just after PD medication}, and HCs who did not explicitly report \textit{not taking any PD medications}).
Finally, to align age distributions across cohorts, we retained subjects aged between 34 and 80 years, corresponding to the full age range of the smallest dataset (\textit{Minsk}).
The resulting cohort composition after all filtering steps is summarized in Table~\ref{tab:patient_counts}, highlighting substantial differences in cohort size, class balance, and gender distribution across datasets.


\section{Methodology}

\subsection{Pre-processing strategy }
To account for heterogeneous acquisition conditions across datasets, all recordings were harmonised through a common pre-processing strategy. Audio signals were peak-normalised, resampled to a common 8\,kHz sampling rate, and trimmed using an energy-based voice activity detector to remove leading and trailing silence.
Recordings were segmented into 50\% overlapping fixed-length windows of two durations, 2.0\,s (without padding) and 4.0\,s (zero-padded), to assess the sensitivity of the proposed framework to temporal context. Padding was applied only for 4.0\,s to avoid excessive data loss ($\sim$80\%) in the smallest external cohorts. 
To mitigate recording-level amplitude differences and prevent models from exploiting trivial energy cues, we applied dataset-level RMS-level equalization based on the training set level distribution. From each segment, we extracted complementary acoustic representations: log--Mel spectrograms, MFCCs, and eGeMAPSv02 features, ensuring consistency with prior work in voice-based neurodegenerative disease detection~\cite{rahmatallah2025pre}.
Further implementation details and parameter settings are provided in the Supplementary Material~(see Section A.2).

\begin{figure*}
  \centering
  \includegraphics[width=0.83\textwidth]{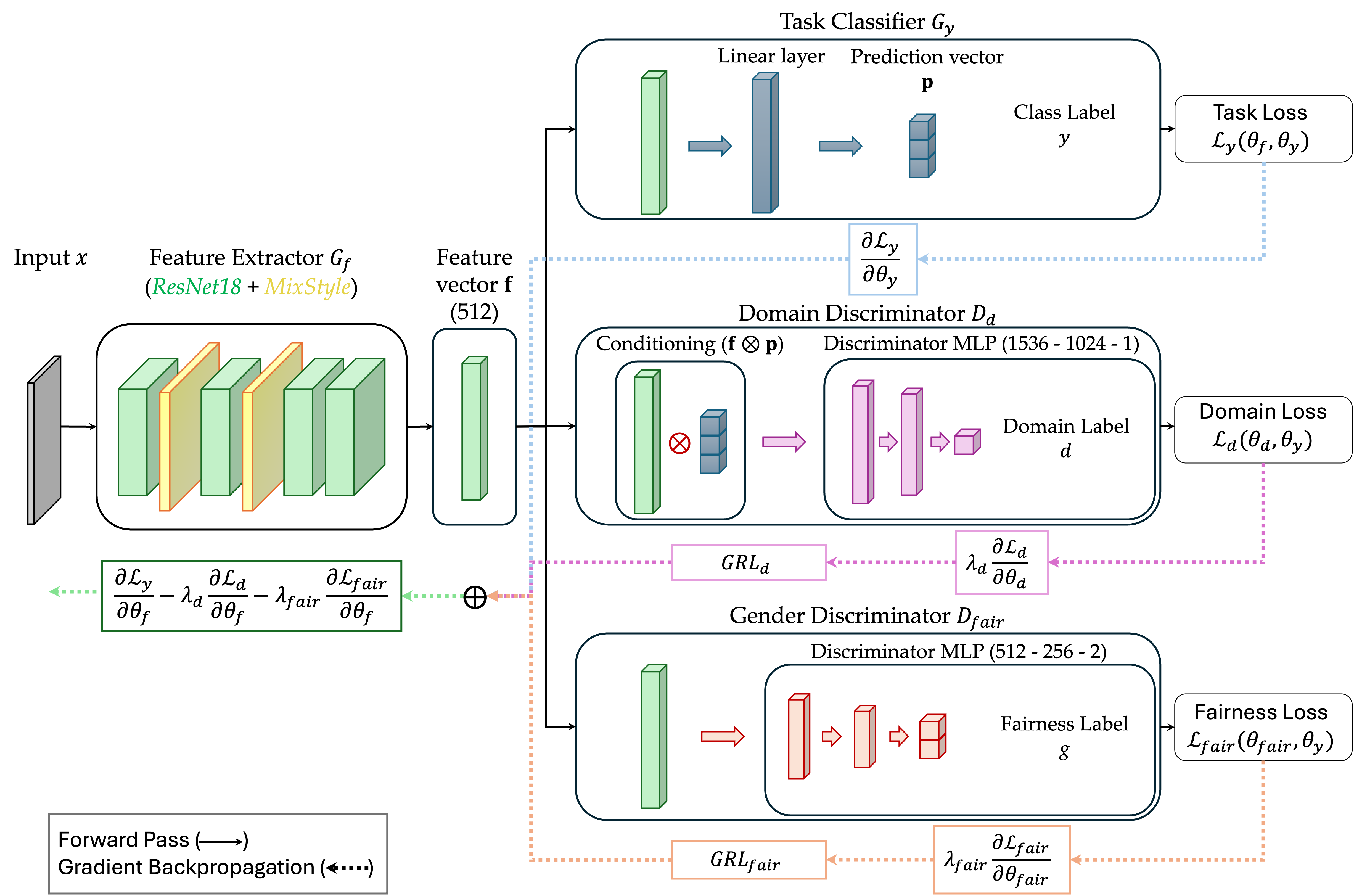}
  \caption{\textit{FairPDA} architecture: ResNet-18+MixStyle extracts a 512-D feature vector; a task head predicts HC/PD/ALS; a conditional domain adversary (input \(f\otimes p\), GRL) performs partial-label alignment via target-driven source reweighting; a GRL-based gender adversary promotes gender-invariant features.}
  \label{fig:model}
\end{figure*}

\subsection{Our Method}
We propose \textit{FairPDA}, a hybrid architecture that integrates domain generalization via style augmentation, partial-label domain adaptation, and adversarial debiasing for fairness. As illustrated in Figure~\ref{fig:model}, the model is built upon a ResNet-18 backbone adapted for single-channel spectrogram input. It maps an input spectrogram $x$ to a feature vector $\mathbf{f} \in \mathbb{R}^{512}$ via a feature extractor $G_f$. To improve robustness to domain shifts, MixStyle~\cite{zhou2021domain} layers are inserted after the first and second residual blocks.
The feature vector $\mathbf{f}$ serves as input to three distinct branches, each associated with a specific learning objective:
\begin{enumerate}
  \item \textit{Task classifier}: A linear classifier $G_y$ maps $\mathbf{f}$ to class probabilities $\mathbf{p}=\text{softmax}(G_y(\mathbf{f}))\in\mathbb{R}^K$, with $K=3$~(HC, PD, ALS). The classifier is trained using the patient-normalized cross-entropy loss $\mathcal{L}_{y}$ described in Section~\ref{sec:experimental_setup}, which mitigates class imbalance and variability in the number of segments per patient.
  \item \textit{Domain Discriminator}: Let $\mathcal{S}$ and $\mathcal{T}$ denote the source and target domain distributions, respectively. To align these domains while accounting for partial-label shifts arising from non-overlapping PD and ALS cohorts, we adopt a Partial CDAN strategy~\cite{cao2018partial}, which selectively aligns shared classes and mitigates negative transfer. Unlike standard DANN, the domain discriminator $D_d$ is conditioned on both feature representations and classifier predictions to preserve class-conditional structure. Specifically, $D_d$ receives as input the multilinear map $\mathbf{h}=\mathbf{f}\otimes\mathbf{p}\in\mathbb{R}^{512\times K}$.
  To reduce the influence of source-only classes, each source sample is weighted by an importance factor $\gamma$, computed as the average predicted probability of the corresponding class $y_s$ over the current target mini-batch $\mathcal{B}_t$:
    \begin{equation}
    \gamma(y_s)=\frac{1}{|\mathcal{B}_t|}\sum_{x_t\in\mathcal{B}_t}p(y=y_s\mid x_t).
    \end{equation}
    The domain adversarial loss $\mathcal{L}_{\textit{d}}$ is defined as a weighted binary cross-entropy:
    \begin{equation}
    \begin{split}
    \mathcal{L}_{\textit{d}} = &- \mathbb{E}_{(x_s, y_s) \sim \mathcal{S}} [\gamma(y_s) \log D_d(\mathbf{f}_s \otimes \mathbf{p}_s)] \\
    &- \mathbb{E}_{x_t \sim \mathcal{T}} [\log (1 - D_d(\mathbf{f}_t \otimes \mathbf{p}_t))]
    \end{split}
    \end{equation}
    where $\mathbb{E}$ denotes the mathematical expectation. 
  \item \textit{Gender discriminator}: To promote demographic invariance, we introduce a fairness branch $D_{\textit{fair}}$, implemented as a 3-layer Multi-Layer Perceptron predicting the binary gender label $g$ from $\mathbf{f}$. The gender discriminator and feature extractor are trained adversarially: while $D_{\textit{fair}}$ aims to correctly predict gender, the feature extractor is encouraged to suppress gender-related information in the shared representation.  The fairness loss is computed as:
  \begin{equation}
    \mathcal{L}_{\textit{fair}} = \mathbb{E}_{(x, g)} [\mathcal{L}_{\text{CE}}(D_{\textit{fair}}(\mathbf{f}), g)]
    \end{equation}
    where $\mathcal{L}_{\text{CE}}$ is the standard cross-entropy loss function.
\end{enumerate}

Although each branch is associated with its own loss function, the overall training dynamics are governed by the shared feature extractor. Let $\theta_f, \theta_y, \theta_d$, and $\theta_{\textit{fair}}$ denote the parameters of the feature extractor $G_f$, task classifier $G_y$, domain discriminator $D_d$, and gender discriminator $D_{\textit{fair}}$, respectively. The total error function $E$ to be optimised is defined as:
\begin{equation}
E(\theta_f, \theta_y, \theta_d, \theta_{\textit{fair}}) = \mathcal{L}_{\textit{y}} - \lambda_{\textit{d}} \mathcal{L}_{\textit{d}} - \lambda_{\textit{fair}} \mathcal{L}_{\textit{fair}}
\end{equation}
Adversarial objectives are implemented via gradient reversal layers (GRLs~\cite{ganin2016domain}) placed between the feature extractor and the discriminators. During backpropagation, GRLs reverse and scale gradients by factors $-\lambda_{\textit{d}}$ and $-\lambda_{\textit{fair}}$, controlling the strength of domain and gender invariance enforced on the shared representation, whereas the overall training implicitly solves a min-max problem via standard gradient-based optimization. Both adversarial weights are linearly warmed up to their final values (see Supplementary Material, Section~A.4). 


\section{Experimental Setup}\label{sec:experimental_setup}
In all experiments, mPower and VOC-ALS are treated as source domains, as they are the largest cohorts and jointly cover the 3-class task, while UAMS and Minsk are used as external target cohorts. In the DG setting, target data are never observed during training, whereas in the UDA setting, 30\% of patients from each target cohort are used as unlabelled adaptation data and the remaining 70\% for external evaluation.
We perform 5-fold cross-validation (CV) on the combined mPower and VOC-ALS cohorts at the patient level.

\textbf{Competitors}\label{subsec:competitors}
We compare \textit{FairPDA} against 12 state-of-the-art models drawn from the PD and ALS voice-analysis literature and from general domain adaptation research (detailed description provided in Supplementary Table~S2), together with three targeted ablations of our method. Baselines include standard ML classifiers, namely Support Vector Machine (SVM) and Gradient Boosted Trees (XGBoost), using MFCCs and eGeMAPSv02 features, as well as DL models operating on spectrograms or raw audio, namely ResNet-18~\cite{rahmatallah2025pre}, ECAPA--TDNN~\cite{desplanques2020ecapa}, and Wav2Vec~2.0~\cite{schneider2019wav2vec}. To assess robustness to domain shift, we evaluate DG methods (MixStyle~\cite{zhou2021domain}, CORAL~\cite{sun2016deep}) and UDA approaches (DANN~\cite{ganin2016domain}, CDAN~\cite{long2018conditional}, and Partial CDAN~\cite{cao2018partial}). We also report three ablations of the proposed model, obtained by removing adversarial warm-up, MixStyle layers, or the gender adversarial branch.

\textbf{Loss functions and optimization}
Our default training objective is the cross-entropy with patient normalization (CE+PN), a re-weighted variant of the standard cross-entropy that enforces equal contribution from each patient. Specifically, each segment-level loss is weighted by \(1/n_i\), where \(n_i\) denotes the number of segments associated with patient \(i\). This prevents patients with a larger number of segments from dominating the optimization process. For comparison, we also evaluate standard cross-entropy (CE). For UDA-based methods, adversarial losses are optimized via GRLs, with linear warm-up schedules used to progressively introduce adversarial objectives and stabilize training. Further implementation details are provided in the Supplementary Material (Section~A.4).

\textbf{Evaluation metrics}
Performance is evaluated at the patient level using Balanced Accuracy (BalAcc), Matthews Correlation Coefficient (MCC), and macro-\(F_1\), all computed at the patient level via soft voting (mean pooling over a patient’s segments). BalAcc and macro-\(F_1\) are reported in \([0,100]\), whilst MCC ranges in \([-1,1]\). In fairness-aware experiments, we additionally report the gender-based Equal Opportunity Difference (EOD) and Equalized Odds Gap (EOG), both computed on the external evaluation cohorts, with lower values indicating better fairness. EOD is defined as the absolute difference between the maximum and minimum true-positive rates across gender groups, whilst EOG corresponds to the larger of the absolute differences in true-positive rates and false-positive rates across gender groups.
We report metrics on the internal test folds of the 5-fold CV and on external cohorts (UAMS and Minsk).


\section{Results}
\begin{table*}[!t]
\caption{Comparison of CE+PN vs.\ CE results for 2.0\,s and 4.0\,s segments. BalAcc is reported as percentage \([0,100]\), whilst MCC ranges from \(-1\) to \(+1\).
Bold: best results for each metric; Orange shading: statistically different from \textit{FairPDA} at $p<0.1$; 
Blue shading: statistically different from \textit{FairPDA} at $p<0.05$.}

\label{tab:balacc_mcc_fairness_cepn_vs_ce_20s_40s}
\centering
\resizebox{\textwidth}{!}{
\begin{tabular}{c|l|l|l|l|l|l|l||l|l|l|l|l|l}
\toprule
\multirow{2}{*}{Seg.} & \multirow{2}{*}{Methods} &
\multicolumn{6}{c||}{CE+PN: Cross-Entropy with Patient Normalization} &
\multicolumn{6}{c}{CE: standard Cross-Entropy}\\
\cmidrule(lr){3-8}\cmidrule(lr){9-14}
& &
Int BalAcc & Int MCC & Ext BalAcc & Ext MCC & EOD & EOG &
Int BalAcc & Int MCC & Ext BalAcc & Ext MCC & EOD & EOG \\
\midrule

\multirow{13}{*}{\rotatebox{90}{2.0\,s}}
& \textit{FairPDA}        & \textbf{52.27 ± 1.78} & 0.28 ± 0.03  & \textbf{45.52 ± 6.75} & \textbf{0.18 ± 0.08} & 0.07 ± 0.09 & 0.11 ± 0.09
                      & \textbf{46.99 ± 2.58} & 0.22 ± 0.06  & \textbf{43.13 ± 2.44} & \textbf{0.14 ± 0.09} & \textbf{0.06 ± 0.04} & \textbf{0.09 ± 0.02} \\
& SVM (MFCC)           & \cellcolor{blue!15}51.37 ± 3.72 & \cellcolor{blue!15}0.21 ± 0.04  & \cellcolor{blue!15}38.54 ± 3.06 & \cellcolor{blue!15}0.01 ± 0.03         & \cellcolor{blue!15}0.06 ± 0.08 & \cellcolor{blue!15}0.13 ± 0.15
                      & \cellcolor{blue!15}41.67 ± 1.91 & \cellcolor{blue!15}0.22 ± 0.05  & \cellcolor{blue!15}33.73 ± 1.62 & \cellcolor{blue!15}-0.01 ± 0.04 & \cellcolor{blue!15}0.17 ± 0.10 & \cellcolor{blue!15}0.18 ± 0.09 \\
& SVM (eG)             & \cellcolor{blue!15}41.46 ± 1.17 & \cellcolor{blue!15}0.23 ± 0.02  & \cellcolor{blue!15}33.26 ± 1.16 & \cellcolor{blue!15}0.02 ± 0.05         & \cellcolor{blue!15}\textbf{0.05 ± 0.03} & \cellcolor{blue!15}\textbf{0.08 ± 0.02}
                      & \cellcolor{blue!15}41.65 ± 1.92 & \cellcolor{blue!15}0.23 ± 0.05  & \cellcolor{blue!15}33.65 ± 1.74 & \cellcolor{blue!15}-0.00 ± 0.03 & \cellcolor{blue!15}0.19 ± 0.07 & \cellcolor{blue!15}0.20 ± 0.08 \\
& XGBoost (MFCC)       & \cellcolor{blue!15}44.34 ± 2.42 & \cellcolor{blue!15}0.22 ± 0.07 & \cellcolor{blue!15}31.68 ± 2.11 & \cellcolor{blue!15}-0.02 ± 0.05        & \cellcolor{blue!15}0.12 ± 0.10 & \cellcolor{blue!15}0.21 ± 0.14
                      & \cellcolor{blue!15}43.90 ± 2.45 & \cellcolor{blue!15}0.22 ± 0.03  & \cellcolor{blue!15}35.12 ± 1.90 & \cellcolor{blue!15}0.04 ± 0.04         & \cellcolor{blue!15}0.20 ± 0.15 & \cellcolor{blue!15}0.22 ± 0.12 \\
& XGBoost (eG)         & \cellcolor{blue!15}45.78 ± 1.58 & \cellcolor{blue!15}0.24 ± 0.07 & \cellcolor{blue!15}28.74 ± 2.13 & \cellcolor{blue!15}-0.05 ± 0.03        & \cellcolor{blue!15}0.06 ± 0.06 & \cellcolor{blue!15}0.19 ± 0.14
                      & \cellcolor{orange!20}46.53 ± 1.64 & \cellcolor{orange!20}0.32 ± 0.03  & \cellcolor{orange!20}38.54 ± 3.06 & \cellcolor{orange!20}0.11 ± 0.07         &\cellcolor{orange!20}0.17 ± 0.05 & \cellcolor{orange!20}0.18 ± 0.05 \\
& ResNet-18            & 50.81 ± 1.75 & 0.29 ± 0.04  & 43.13 ± 6.25 & 0.13 ± 0.08          & 0.07 ± 0.07 & 0.15 ± 0.07
                      & \cellcolor{orange!20}45.66 ± 5.79 & \cellcolor{orange!20}0.26 ± 0.05  & \cellcolor{orange!20}42.78 ± 5.96 & \cellcolor{orange!20}0.11 ± 0.07          & \cellcolor{orange!20}0.15 ± 0.11 & \cellcolor{orange!20}0.20 ± 0.09 \\
& ECAPA--TDNN          & \cellcolor{blue!15}51.37 ± 3.72 & \cellcolor{blue!15}0.28 ± 0.07  & \cellcolor{blue!15}32.60 ± 4.62 & \cellcolor{blue!15}-0.00 ± 0.06        & \cellcolor{blue!15}0.08 ± 0.07 & \cellcolor{blue!15}0.15 ± 0.06
                      & \cellcolor{blue!15}46.52 ± 4.75 & \cellcolor{blue!15}0.24 ± 0.06  & \cellcolor{blue!15}35.35 ± 4.09 & \cellcolor{blue!15}0.03 ± 0.06         & \cellcolor{blue!15}0.08 ± 0.05 & \cellcolor{blue!15}0.15 ± 0.07 \\
& Wav2Vec 2.0          & \cellcolor{orange!20}44.77 ± 6.28 & \cellcolor{orange!20}0.14 ± 0.04 & \cellcolor{orange!20}38.80 ± 0.29 & \cellcolor{orange!20}0.11 ± 0.10          & \cellcolor{orange!20}0.10 ± 0.08 & \cellcolor{orange!20}0.10 ± 0.08
                      & \cellcolor{blue!15}37.56 ± 2.08 & \cellcolor{blue!15}0.14 ± 0.04 & \cellcolor{blue!15}39.37 ± 5.84 & \cellcolor{blue!15}0.09 ± 0.08          & \cellcolor{blue!15}0.11 ± 0.08 & \cellcolor{blue!15}0.11 ± 0.08 \\
& MixStyle             & \cellcolor{blue!15}42.29 ± 2.81 & \cellcolor{blue!15}0.22 ± 0.03  & \cellcolor{blue!15}39.67 ± 7.19 & \cellcolor{blue!15}0.11 ± 0.12          & \cellcolor{blue!15}0.10 ± 0.08 & \cellcolor{blue!15}0.13 ± 0.05
                      & 44.02 ± 3.23 & 0.21 ± 0.02  & 42.52 ± 6.53 & 0.13 ± 0.04          & 0.14 ± 0.11 & 0.21 ± 0.13 \\
& CORAL                & \cellcolor{blue!15}45.97 ± 5.30 & \cellcolor{blue!15}\textbf{0.36 ± 0.07} & \cellcolor{blue!15}30.48 ± 5.09 & \cellcolor{blue!15}-0.01 ± 0.08        & \cellcolor{blue!15}0.19 ± 0.08 & \cellcolor{blue!15}0.25 ± 0.09
                      & \cellcolor{blue!15}43.79 ± 4.92 & \cellcolor{blue!15}\textbf{0.38 ± 0.07} & \cellcolor{blue!15}34.93 ± 6.49 & \cellcolor{blue!15}0.05 ± 0.10         & \cellcolor{blue!15}0.09 ± 0.06 & \cellcolor{blue!15}0.15 ± 0.08 \\
& DANN                 & \cellcolor{blue!15}49.82 ± 4.48 & \cellcolor{blue!15}0.26 ± 0.04  & \cellcolor{blue!15}38.10 ± 4.14 & \cellcolor{blue!15}0.08 ± 0.05          & \cellcolor{blue!15}0.15 ± 0.10 & \cellcolor{blue!15}0.15 ± 0.10
                      & \cellcolor{orange!20}45.42 ± 6.38 & \cellcolor{orange!20}0.25 ± 0.05  & \cellcolor{orange!20}41.55 ± 5.12 & \cellcolor{orange!20}0.12 ± 0.08         & \cellcolor{orange!20}0.12 ± 0.11 & \cellcolor{orange!20}0.20 ± 0.05 \\
& CDAN                 & 45.53 ± 11.07 & 0.23 ± 0.10 & 41.10 ± 5.20 & 0.12 ± 0.07          & 0.09 ± 0.04 & 0.11 ± 0.04
                      & \cellcolor{blue!15}40.38 ± 6.90 & \cellcolor{blue!15}0.18 ± 0.06  & \cellcolor{blue!15}33.97 ± 4.94 & \cellcolor{blue!15}0.03 ± 0.07         & \cellcolor{blue!15}0.15 ± 0.11 & \cellcolor{blue!15}0.24 ± 0.17 \\
& Partial CDAN         & 48.50 ± 6.34 & 0.28 ± 0.05  & 39.07 ± 10.69 & 0.10 ± 0.15         & 0.09 ± 0.04 & 0.20 ± 0.14
                      & 42.68 ± 3.23 & 0.20 ± 0.10  & 41.56 ± 6.82 & 0.12 ± 0.09         & 0.14 ± 0.11 & 0.20 ± 0.11 \\

\midrule

\multirow{13}{*}{\rotatebox{90}{4.0\,s}}
& \textit{FairPDA}        & 50.91 ± 7.77 & 0.26 ± 0.09  & \textbf{42.19 ± 2.06} & \textbf{0.19 ± 0.06} & \textbf{0.04 ± 0.12} & \textbf{0.06 ± 0.04}
                      & 50.37 ± 4.26 & 0.29 ± 0.06  & \textbf{45.87 ± 4.88} & \textbf{0.17 ± 0.07} & \textbf{0.04 ± 0.01} & \textbf{0.07 ± 0.06} \\
& SVM (MFCC)           & \cellcolor{blue!15}42.10 ± 7.60 & \cellcolor{blue!15}0.24 ± 0.04  & \cellcolor{blue!15}37.57 ± 9.37 & \cellcolor{blue!15}0.09 ± 0.08          & \cellcolor{blue!15}0.06 ± 0.08 & \cellcolor{blue!15}0.13 ± 0.26
                      & \cellcolor{blue!15}43.67 ± 3.24 & \cellcolor{blue!15}0.23 ± 0.08 & \cellcolor{blue!15}38.19 ± 2.54 & \cellcolor{blue!15}0.09 ± 0.15          & \cellcolor{blue!15}0.25 ± 0.09 & \cellcolor{blue!15}0.23 ± 0.09 \\
& SVM (eG)             & \cellcolor{blue!15}51.56 ± 3.14 & \cellcolor{blue!15}0.24 ± 0.04  & \cellcolor{blue!15}40.20 ± 3.82 & \cellcolor{blue!15}0.09 ± 0.03          & \cellcolor{blue!15}0.05 ± 0.06 & \cellcolor{blue!15}0.12 ± 0.04
                      & \cellcolor{blue!15}42.52 ± 2.59 & \cellcolor{blue!15}0.24 ± 0.06  & \cellcolor{blue!15}37.37 ± 0.74 & \cellcolor{blue!15}0.09 ± 0.13 & \cellcolor{blue!15}0.22 ± 0.07 & \cellcolor{blue!15}0.22 ± 0.07 \\
& XGBoost (MFCC)       & \cellcolor{blue!15}49.35 ± 2.84 & \cellcolor{blue!15}0.22 ± 0.07 & \cellcolor{blue!15}31.68 ± 2.11 & \cellcolor{blue!15}-0.02 ± 0.05        & \cellcolor{blue!15}0.12 ± 0.10 & \cellcolor{blue!15}0.21 ± 0.14
                      & \cellcolor{blue!15}45.00 ± 1.92 & \cellcolor{blue!15}0.23 ± 0.02 & \cellcolor{blue!15}32.86 ± 2.05 & \cellcolor{blue!15}-0.00 ± 0.05        & \cellcolor{blue!15}0.12 ± 0.10 & \cellcolor{blue!15}0.13 ± 0.09 \\
& XGBoost (eG)         & \cellcolor{blue!15}52.38 ± 9.28 & \cellcolor{blue!15}0.24 ± 0.07 & \cellcolor{blue!15}34.82 ± 2.94 & \cellcolor{blue!15}-0.05 ± 0.03        & \cellcolor{blue!15}0.06 ± 0.06 & \cellcolor{blue!15}0.19 ± 0.14
                      & \cellcolor{blue!15}46.57 ± 1.08 & \cellcolor{blue!15}0.32 ± 0.03 & \cellcolor{blue!15}37.52 ± 1.87 & \cellcolor{blue!15}0.09 ± 0.05          & \cellcolor{blue!15}0.12 ± 0.05 & \cellcolor{blue!15}0.17 ± 0.04 \\
& ResNet-18            & \cellcolor{orange!20}46.35 ± 6.80 & \cellcolor{orange!20}0.31 ± 0.08  & \cellcolor{orange!20}39.85 ± 8.24 & \cellcolor{orange!20}0.12 ± 0.03          & \cellcolor{orange!20}0.09 ± 0.08 & \cellcolor{orange!20}0.12 ± 0.05
                      & \cellcolor{orange!20}48.21 ± 6.68 & \cellcolor{orange!20}0.32 ± 0.05  & \cellcolor{orange!20}43.64 ± 4.59 & \cellcolor{orange!20}0.12 ± 0.04          & \cellcolor{orange!20}0.10 ± 0.08 & \cellcolor{orange!20}0.13 ± 0.05 \\
& ECAPA--TDNN          & \cellcolor{blue!15}48.87 ± 5.47 & \cellcolor{blue!15}0.28 ± 0.08  & \cellcolor{blue!15}33.19 ± 5.59 & \cellcolor{blue!15}-0.00 ± 0.06        & \cellcolor{blue!15}0.11 ± 0.11 & \cellcolor{blue!15}0.15 ± 0.14
                      & \cellcolor{blue!15}51.68 ± 3.32 & \cellcolor{blue!15}0.28 ± 0.03  & \cellcolor{blue!15}32.39 ± 3.72 & \cellcolor{blue!15}-0.01 ± 0.04        & \cellcolor{blue!15}0.14 ± 0.10 & \cellcolor{blue!15}0.16 ± 0.08 \\
& Wav2Vec 2.0          & \cellcolor{blue!15}32.85 ± 4.00 & \cellcolor{blue!15}0.09 ± 0.06 & \cellcolor{blue!15}38.28 ± 3.75 & \cellcolor{blue!15}0.08 ± 0.06         & \cellcolor{blue!15}0.06 ± 0.04 & \cellcolor{blue!15}0.09 ± 0.12
                      & \cellcolor{blue!15}33.53 ± 3.73 & \cellcolor{blue!15}0.10 ± 0.06 & \cellcolor{blue!15}38.45 ± 3.13 & \cellcolor{blue!15}0.09 ± 0.16         & \cellcolor{blue!15}0.14 ± 0.04 & \cellcolor{blue!15}0.16 ± 0.08 \\
& MixStyle             & \cellcolor{orange!20}51.66 ± 4.04 & \cellcolor{orange!20}0.30 ± 0.03  & \cellcolor{orange!20}39.04 ± 6.27 & \cellcolor{orange!20}0.08 ± 0.09         & \cellcolor{orange!20}0.12 ± 0.08 & \cellcolor{orange!20}0.14 ± 0.07
                      & \cellcolor{orange!20}51.20 ± 5.06 & \cellcolor{orange!20}0.26 ± 0.04  & \cellcolor{orange!20}40.63 ± 4.73 & \cellcolor{orange!20}0.14 ± 0.07          & \cellcolor{orange!20}0.12 ± 0.07 & \cellcolor{orange!20}0.14 ± 0.07 \\
& CORAL                & \cellcolor{blue!15}\textbf{54.62 ± 5.96} & \cellcolor{blue!15}\textbf{0.36 ± 0.06} & \cellcolor{blue!15}36.32 ± 3.90 & \cellcolor{blue!15}0.06 ± 0.06         & \cellcolor{blue!15}0.13 ± 0.09 & \cellcolor{blue!15}0.13 ± 0.09
                      & \cellcolor{blue!15}52.09 ± 1.35 & \cellcolor{blue!15}\textbf{0.35 ± 0.06} & \cellcolor{blue!15}36.93 ± 6.24 & \cellcolor{blue!15}0.07 ± 0.09         & \cellcolor{blue!15}0.14 ± 0.08 & \cellcolor{blue!15}0.21 ± 0.09 \\
& DANN                 & 46.20 ± 5.90 & 0.26 ± 0.07  & 39.86 ± 8.47 & 0.10 ± 0.10         & 0.17 ± 0.08 & 0.17 ± 0.08
                      & \cellcolor{orange!20}47.34 ± 2.99 & \cellcolor{orange!20}0.25 ± 0.11  & \cellcolor{orange!20}37.91 ± 8.31 & \cellcolor{orange!20}0.11 ± 0.13         & \cellcolor{orange!20}0.14 ± 0.13 & \cellcolor{orange!20}0.27 ± 0.04 \\
& CDAN                 & \cellcolor{blue!15}51.09 ± 2.56 & \cellcolor{blue!15}0.27 ± 0.05  & \cellcolor{blue!15}38.33 ± 1.34 & \cellcolor{blue!15}0.09 ± 0.04         & \cellcolor{blue!15}0.20 ± 0.07 & \cellcolor{blue!15}0.22 ± 0.10
                      & 49.16 ± 5.56 & 0.29 ± 0.06  & 44.33 ± 6.55 & 0.16 ± 0.10 & 0.11 ± 0.08 & 0.16 ± 0.10 \\
& Partial CDAN         & \cellcolor{blue!15}48.68 ± 7.80 & \cellcolor{blue!15}0.28 ± 0.06  & \cellcolor{blue!15}36.10 ± 5.70 & \cellcolor{blue!15}0.04 ± 0.07         & \cellcolor{blue!15}0.22 ± 0.07 & \cellcolor{blue!15}0.24 ± 0.07
                      & \cellcolor{blue!15}\textbf{53.79 ± 4.92} & \cellcolor{blue!15}0.28 ± 0.05  & \cellcolor{blue!15}34.93 ± 6.49 & \cellcolor{blue!15}0.06 ± 0.04         & \cellcolor{blue!15}0.12 ± 0.08 & \cellcolor{blue!15}0.17 ± 0.06 \\

\bottomrule
\end{tabular}}
\end{table*}

Table~\ref{tab:balacc_mcc_fairness_cepn_vs_ce_20s_40s} summarizes patient-level internal/external performance (BalAcc, MCC) and fairness (EOD, EOG) for \textit{FairPDA} and all competitors. Macro-\(F_1\) is reported in Supplementary Tables~S3--S4. Across all tables, shading indicates methods significantly different from \textit{FairPDA} on external predictions (paired test; orange $p<0.1$, blue $p<0.05$; see Supplementary Section~B for further details).

It is worth noting that the proposed approach achieves the highest External MCC and External BalAcc across both 2.0\,s and 4.0\,s segments under CE and CE+PN (Table~\ref{tab:balacc_mcc_fairness_cepn_vs_ce_20s_40s}). 
In addition, \textit{FairPDA} attains the lowest gender gaps for 2.0\,s with CE, 4.0\,s with CE+PN, and 4.0\,s with CE.

\begin{table*}[!t]
\caption{Ablation study comparing CE+PN vs.\ CE for 2.0\,s and 4.0\,s segments. BalAcc is reported as percentage \([0,100]\), whilst MCC ranges from \(-1\) to \(+1\).
Bold: best results for each metric; Orange shading: statistically different from \textit{FairPDA} at $p<0.1$; 
Blue shading: statistically different from \textit{FairPDA} at $p<0.05$.}
\label{tab:abl_cepn_vs_ce_balacc_mcc_fairness}
\centering
\resizebox{\textwidth}{!}{
\begin{tabular}{c|l|l|l|l|l|l|l||l|l|l|l|l|l}
\toprule
\multirow{2}{*}{Seg.} &
\multicolumn{7}{c||}{CE+PN: Cross-Entropy with Patient Normalization} &
\multicolumn{6}{c}{CE: standard Cross-Entropy} \\
\cmidrule(lr){2-8}\cmidrule(lr){9-14}
& Method & Int BalAcc & Int MCC & Ext BalAcc & Ext MCC & EOD & EOG
& Int BalAcc & Int MCC & Ext BalAcc & Ext MCC & EOD & EOG \\
\midrule

\multirow{4}{*}{\rotatebox{90}{2.0\,s}}
& \textit{FairPDA}        & \textbf{52.27 ± 1.78} & 0.28 ± 0.03  & \textbf{45.52 ± 6.75} & \textbf{0.18 ± 0.08} & 0.07 ± 0.09 & 0.11 ± 0.09
& \textbf{46.99 ± 2.58} & 0.22 ± 0.06  & \textbf{43.13 ± 2.44} & \textbf{0.14 ± 0.09} & \textbf{0.06 ± 0.04} & \textbf{0.09 ± 0.02} \\
& no Warmup
& \cellcolor{blue!15}46.04 ± 9.57 & \cellcolor{blue!15}0.21 ± 0.07 & \cellcolor{blue!15}39.73 ± 4.83 & \cellcolor{blue!15}0.10 ± 0.07 & \cellcolor{blue!15}0.13 ± 0.08 & \cellcolor{blue!15}0.20 ± 0.15
& \cellcolor{blue!15}44.73 ± 9.16 & \cellcolor{blue!15}0.23 ± 0.04 & \cellcolor{blue!15}37.69 ± 4.70 & \cellcolor{blue!15}0.08 ± 0.06 & \cellcolor{blue!15}0.13 ± 0.09 & \cellcolor{blue!15}0.21 ± 0.13 \\
& no MixStyle
& \cellcolor{blue!15}46.06 ± 4.92 & \cellcolor{blue!15}0.26 ± 0.05 & \cellcolor{blue!15}39.12 ± 4.72 & \cellcolor{blue!15}0.08 ± 0.08 & \cellcolor{blue!15}\textbf{0.21 ± 0.10} & \cellcolor{blue!15}\textbf{0.23 ± 0.08}
& \cellcolor{blue!15}45.50 ± 3.89 & \cellcolor{blue!15}0.23 ± 0.06 & \cellcolor{blue!15}38.76 ± 4.10 & \cellcolor{blue!15}0.09 ± 0.06 & \cellcolor{blue!15}0.16 ± 0.07 & \cellcolor{blue!15}0.25 ± 0.09 \\
& no Fairness
& \cellcolor{blue!15}46.23 ± 5.79 & \cellcolor{blue!15}\textbf{0.39 ± 0.06} & \cellcolor{blue!15}34.54 ± 5.21 & \cellcolor{blue!15}0.04 ± 0.07 & \cellcolor{blue!15}0.13 ± 0.15 & \cellcolor{blue!15}0.14 ± 0.15
& \cellcolor{blue!15}39.95 ± 4.84 & \cellcolor{blue!15}\textbf{0.38 ± 0.03} & \cellcolor{blue!15}37.78 ± 12.26 & \cellcolor{blue!15}0.06 ± 0.02 & \cellcolor{blue!15}0.23 ± 0.06 & \cellcolor{blue!15}0.23 ± 0.06 \\
\midrule

\multirow{4}{*}{\rotatebox{90}{4.0\,s}}
& \textit{FairPDA}        & \textbf{50.91 ± 7.77} & 0.26 ± 0.09  & \textbf{42.19 ± 2.06} & \textbf{0.19 ± 0.06} & \textbf{0.04 ± 0.12} & \textbf{0.06 ± 0.04}
                      & \textbf{50.37 ± 4.26} & 0.29 ± 0.06  & \textbf{45.87 ± 4.88} & \textbf{0.17 ± 0.07} & \textbf{0.04 ± 0.01} & \textbf{0.07 ± 0.06} \\
& no Warmup
& \cellcolor{blue!15}50.33 ± 5.19 & \cellcolor{blue!15}\textbf{0.31 ± 0.05} & \cellcolor{blue!15}38.79 ± 4.13 & \cellcolor{blue!15}0.10 ± 0.06 & \cellcolor{blue!15}0.14 ± 0.13 & \cellcolor{blue!15}0.22 ± 0.06
& 41.37 ± 9.55 & \textbf{0.17 ± 0.09} & 39.32 ± 7.19 & 0.09 ± 0.11 & 0.10 ± 0.10 & 0.15 ± 0.10 \\
& no MixStyle
& \cellcolor{orange!20}47.77 ± 8.53 & \cellcolor{orange!20}0.28 ± 0.07 & \cellcolor{orange!20}38.19 ± 12.05 & \cellcolor{orange!20}0.07 ± 0.17 & \cellcolor{orange!20}0.22 ± 0.01 & \cellcolor{orange!20}0.22 ± 0.01
& \cellcolor{orange!20}46.42 ± 9.73 & \cellcolor{orange!20}0.25 ± 0.08 & \cellcolor{orange!20}36.67 ± 7.22 & \cellcolor{orange!20}0.10 ± 0.07 & \cellcolor{orange!20}0.14 ± 0.10 & \cellcolor{orange!20}0.18 ± 0.08 \\
& no Fairness
& \cellcolor{blue!15}50.74 ± 11.57 & \cellcolor{blue!15}0.29 ± 0.09 & \cellcolor{blue!15}37.27 ± 6.69 & \cellcolor{blue!15}0.06 ± 0.12 & \cellcolor{blue!15}0.14 ± 0.18 & \cellcolor{blue!15}0.23 ± 0.05
& \cellcolor{blue!15}47.39 ± 4.37 & \cellcolor{blue!15}0.25 ± 0.06 & \cellcolor{blue!15}35.33 ± 9.80 & \cellcolor{blue!15}0.04 ± 0.14 & \cellcolor{blue!15}0.19 ± 0.09 & \cellcolor{blue!15}0.23 ± 0.08 \\
\bottomrule
\end{tabular}
}
\end{table*}

\textbf{Internal gains do not translate to external generalization.}
We evaluate 12 competing methods under four configurations (2.0\,s/4.0\,s $\times$ CE+PN/CE), for a total of 48 competitor configurations. 
Paired tests on external predictions indicate statistically significant differences with respect to \textit{FairPDA} in 41 out of these 48 configurations. In 17/48 runs, competitors achieve higher internal discrimination than \textit{FairPDA} (up to Int BalAcc 54.62 and Int MCC 0.38, vs. \textit{FairPDA}'s best internal results of 52.27 and 0.29), but do not preserve this advantage under cross-cohort testing, where Ext MCC is often close to zero and occasionally negative (Table~\ref{tab:balacc_mcc_fairness_cepn_vs_ce_20s_40s}). 
To quantify the internal--external transfer, we report 
$\Delta_{\mathrm{IE}}=\mathrm{MCC}_{\mathrm{Ext}}-\mathrm{MCC}_{\mathrm{Int}}$.
Across configurations, competitors exhibit substantial internal--external drops, with 
$\Delta_{\mathrm{IE}} \in [-0.37, -0.15]$, 
whereas \textit{FairPDA} shows smaller gaps 
($\Delta_{\mathrm{IE}} \in [-0.12, -0.07]$).
We hypothesize that this mismatch may reflect overfitting to cohort-specific properties (e.g., differences in recording conditions or dataset composition), although we do not directly identify such cues in our analysis.
Absolute performance values are moderate across all methods, showcasing a demanding and clinically realistic HC/PD/ALS cross-cohort setting: PD and ALS may exhibit partially overlapping phonatory alterations, and class-discriminative information in sustained vowels lies in subtle acoustic deviations rather than in coarse temporal structure. 

\textbf{Patient normalization improves robustness under cohort imbalance.}
Cohort- and class-level imbalance (Table~\ref{tab:patient_counts}) further increases the difficulty of cross-cohort transfer.
Comparing CE and CE+PN settings (Table~\ref{tab:balacc_mcc_fairness_cepn_vs_ce_20s_40s}), for 2.0\,s segments, CE+PN increases External MCC from 0.14$\pm$0.09 to 0.18$\pm$0.08, whilst for 4.0\,s it improves from 0.17$\pm$0.07 to 0.19$\pm$0.06. 
External BalAcc improves for 2.0\,s (+2.4 points), whilst for 4.0\,s it is lower under CE+PN. 
Patient-level normalization reduces the dominance of larger cohorts and patients contributing many overlapping segments, thereby mitigating reliance on acquisition-specific cues that fail to transfer across cohorts.

\textbf{Fairness: lowest gender gap in most settings.}
Across the four configurations (2.0\,s/4.0\,s $\times$ CE/CE+PN), \textit{FairPDA} achieves the lowest fairness gaps in three out of four cases.
In particular, it attains the minimum EOD/EOG for 2.0\,s with CE (0.06/0.09), 4.0\,s with CE+PN (0.04/0.06), and 4.0\,s with CE (0.04/0.07) (Table~\ref{tab:balacc_mcc_fairness_cepn_vs_ce_20s_40s}).
For 2.0\,s with CE+PN, SVM(eG) attains slightly lower EOD/EOG (0.05/0.08), whereas \textit{FairPDA} remains close (0.07/0.11) and achieves the best External MCC, supporting the most favourable accuracy--fairness trade-off under cross-cohort evaluation.

\textbf{Ablations: Fairness as a targeted regulariser.}
Ablations (Table~\ref{tab:abl_cepn_vs_ce_balacc_mcc_fairness}) support the role of each component in external transfer and fairness. 
Under CE+PN, removing adversarial warm-up reduces External MCC from 0.18$\pm$0.08 to 0.10$\pm$0.07 (2.0\,s) and from 0.19$\pm$0.06 to 0.10$\pm$0.06 (4.0\,s), whereas removing MixStyle further degrades external performance (e.g., to 0.08$\pm$0.08 for 2.0\,s). 
We also noticed that removing the adversarial gender branch increases internal separability but leads to a marked drop in External MCC (0.18$\pm$0.08 to 0.04$\pm$0.07 for 2.0\,s CE+PN) and worsens gender fairness (e.g., EOD 0.07$\pm$0.09 to 0.13$\pm$0.15), indicating that gender invariance acts as a targeted regulariser that improves the discrimination--fairness trade-off under domain shift.

\textbf{Segment length: limited benefit from longer windows.}
No significant external gains are observed when increasing the segment length from 2.0\,s to 4.0\,s. For CE+PN (Table~\ref{tab:balacc_mcc_fairness_cepn_vs_ce_20s_40s}), External MCC improves only marginally (0.18$\pm$0.08 to 0.19$\pm$0.06), whilst External BalAcc slightly decreases (45.52$\pm$6.75 to 42.19$\pm$2.06). Under CE, longer segments yield slightly higher external performance for \textit{FairPDA}, but the observed improvements vary across settings and methods. This suggests that sustained vowels are quasi-stationary signals, such that longer, zero-padded windows mainly replicate similar phonatory patterns rather than introducing additional disease-discriminative information.

\section{Conclusions}
We studied cross-device and cross-cohort HC/PD/ALS voice classification under partial-label mismatch, explicitly evaluating gender fairness. We proposed \textit{FairPDA}, a hybrid framework combining MixStyle augmentation, conditional partial-label adversarial alignment, and an adversarial gender-invariance branch. 
In the cross-cohort evaluation, \textit{FairPDA} achieves the highest mean External MCC and External BalAcc in all four configurations (2.0\,s/4.0\,s $\times$ CE/CE+PN) and attains the lowest gender gaps in three out of four cases (Table~\ref{tab:balacc_mcc_fairness_cepn_vs_ce_20s_40s}). 
Across configurations, paired statistical analyses on external predictions further support the robustness of \textit{FairPDA} under cross-cohort shift.

This work also provides a first benchmark for ternary HC/PD/ALS classification across heterogeneous sustained-vowel cohorts, serving as a concrete testbed for studying learning under partial-label domain shift and demographic bias.
Future work will consider larger, more diverse cohorts and will focus on imbalance-aware objectives with improved mechanisms to detect and down-weight source-only classes under partial-label shift. Finally, our fairness analysis is currently limited to binary gender labels due to metadata availability; extending the evaluation to additional sensitive attributes and intersectional subgroup analyses represents an important direction for future research.

\section*{Acknowledgments}
Arianna Francesconi is a Ph.D. student enrolled in the National Ph.D. in Artificial Intelligence, XXXIX cycle, course on Health and Life Sciences, organized by Università Campus Bio-Medico di Roma.
This work was partially funded by: (i) PNRR – DM 117/2023; (ii) Eustema S.p.A.

\section*{Declaration of competing interest}
The authors declare that they have no known competing financial
interests or personal relationships that could have appeared to
influence the work reported in this paper.

\section*{Data availability}
All data used in this study are publicly available from the sources.
No new data were collected by the authors.

\bibliographystyle{elsarticle-num}
\bibliography{references}

@article{bot2016mpower,
  title={{The mPower study, Parkinson disease mobile data collected using ResearchKit}},
  author={Bot, Brian M and others},
  journal={Scientific data},
  year={2016},
  publisher={Nature Publishing Group}
}

@article{zhou2021domain,
  title={{Domain generalization with mixstyle}},
  author={Zhou, Kaiyang and others},
  journal={arXiv:2104. 02008},
  year={2021}
}

@inproceedings{sun2016deep,
  title={{Deep coral: Correlation alignment for deep domain adaptation}},
  author={Sun, Baochen and Saenko, Kate},
  booktitle={European conference on computer vision},
  year={2016},
  organization={Springer}
}

@article{desplanques2020ecapa,
  title={{Ecapa-tdnn: Emphasized channel attention, propagation and aggregation in tdnn based speaker verification}},
  author={Desplanques, Brecht and others},
  journal={arXiv:2005.07143},
  year={2020}
}

@article{schneider2019wav2vec,
  title={{wav2vec: Unsupervised pre-training for speech recognition}},
  author={Schneider, Steffen and others},
  journal={arXiv:1904.05862},
  year={2019}
}

@article{ganin2016domain,
  title={{Domain-adversarial training of neural networks}},
  author={Ganin, Yaroslav and others},
  journal={Journal of machine learning research},
  year={2016}
}

@article{long2018conditional,
  title={{Conditional adversarial domain adaptation}},
  author={Long, Mingsheng and others},
  journal={Advances in neural information processing systems},
  year={2018}
}

@inproceedings{cao2018partial,
  title={{Partial adversarial domain adaptation}},
  author={Cao, Zhangjie and others},
  booktitle={Proceedings of the European conference on computer vision (ECCV)},
  year={2018}
}

@article{prior2023voice,
  title={{Voice Samples for Patients with Parkinson’s Disease and Healthy Controls}},
  author={Prior, F and others},
  journal={Dataset, figshare},
  year={2023}
}

@article{dubbioso2024voice,
  title={{Voice signals database of ALS patients with different dysarthria severity and healthy controls}},
  author={Dubbioso, Raffaele and others},
  journal={Scientific Data},
  year={2024},
  publisher={Nature Publishing Group UK London}
}

@article{vashkevich2021classification,
  title={{Classification of ALS patients based on acoustic analysis of sustained vowel phonations}},
  author={Vashkevich, Maxim and Rushkevich, Yu},
  journal={Biomedical Signal Processing and Control},
  year={2021},
  publisher={Elsevier}
}

@article{rahmatallah2025pre,
  title={{Pre-trained convolutional neural networks identify Parkinson’s disease from spectrogram images of voice samples}},
  author={Rahmatallah, Yasir and others},
  journal={Scientific Reports},
  year={2025},
  publisher={Nature Publishing Group UK London}
}

@article{eyben2015geneva,
  title={{The Geneva minimalistic acoustic parameter set (GeMAPS) for voice research and affective computing}},
  author={Eyben, Florian and others},
  journal={IEEE Transactions on Affective Computing},
  year={2015},
  publisher={IEEE}
}

@inproceedings{gope2020raw,
  title={{Raw speech waveform based classification of patients with ALS, Parkinson’s disease and healthy controls using CNN-BLSTM}},
  author={Gope, Dipanjan and Ghosh, Prasanta Kumar},
  booktitle={Proc. Interspeech},
  year={2020}
}

@article{sedigh2025voice,
  title={{Voice-Based Detection of Parkinson’s Disease Using Machine and Deep Learning Approaches: A Systematic Review}},
  author={Sedigh Malekroodi, Hadi and others},
  journal={Bioengineering},
  year={2025},
  publisher={MDPI}
}

@article{yang2024deconstructing,
  title={{Deconstructing demographic bias in speech-based machine learning models for digital health},
  author={Yang, Michael and others}},
  journal={Frontiers in Digital Health},
  year={2024},
  publisher={Frontiers Media SA}
}

@article{yang2024limits,
  title={{The limits of fair medical imaging AI in real-world generalization}},
  author={Yang, Yuzhe and others},
  journal={Nature Medicine},
  year={2024},
  publisher={Nature Publishing Group US New York}
}

@article{yang2023adversarial,
  title={{An adversarial training framework for mitigating algorithmic biases in clinical machine learning}},
  author={Yang, Jenny and others},
  journal={NPJ digital medicine},
  year={2023},
  publisher={Nature Publishing Group UK London}
}

@article{ibarra2023towards,
  title={{Towards a corpus (and language)-independent screening of parkinson’s disease from voice and speech through domain adaptation}},
  author={Ibarra, Emiro J and others},
  journal={Bioengineering},
  year={2023},
  publisher={MDPI}
}

@article{xu2022algorithmic,
  title={{Algorithmic fairness in computational medicine}},
  author={Xu, Jie and others},
  journal={EBioMedicine},
  year={2022},
  publisher={Elsevier}
}

@article{guan2021domain,
  title={{Domain adaptation for medical image analysis: a survey}},
  author={Guan, Hao and Liu, Mingxia},
  journal={IEEE Transactions on Biomedical Engineering},
  year={2021},
  publisher={IEEE}
}

@article{zhou2024mixstyle,
  title={{Mixstyle neural networks for domain generalization and adaptation}},
  author={Zhou, Kaiyang and others},
  journal={International Journal of Computer Vision},
  year={2024},
  publisher={Springer}
}

\end{document}